\documentstyle[twocolumn,prl,aps,epsf]{revtex}

\newcommand{\epsplace}[1]{\epsffile{#1}}

\begin{document}


\title{Is   Heteropolymer Freezing  Well Described by the Random
Energy Model?}
\author{Vijay S. Pande\dag, Alexander Yu. Grosberg \dag \ddag,
Chris Joerg\S, Toyoichi Tanaka\dag}

\address{
\dag Department of Physics and Center for Materials Science and
Engineering, \\ Massachusetts Institute of Technology, Cambridge,
Massachusetts 02139,  USA \\
\ddag {\em On leave from:\/} Institute of Chemical Physics, 
Russian Academy of Sciences, Moscow 117977, Russia \\
\S Laboratory for Computer Science,  Massachusetts Institute of
Technology, Cambridge, Massachusetts 02139,  USA
}

\address{
{\em \bigskip
\begin{quote}
It is widely held that the Random Energy Model (REM)  describes
the freezing transition of a variety of types of heteropolymers.
We demonstrate that the hallmark property of REM, statistical
independence of the energies of states over disorder, is violated
in different ways for models commonly employed in heteropolymer
freezing studies. The implications for proteins are also discussed.
\end{quote} } }


\maketitle


\section{Introduction}

Heteropolymer freezing is widely recognized as a model for protein
folding.  By heteropolymer freezing, we refer to the transition 
from a phase in which many conformations (${\cal
O}(e^N)$) dominate equilibrium to one in which only one or very
few (${\cal O}(1)$) are thermodynamically relevant.  Remarkable
progress in the understanding of heteropolymer freezing has been 
achieved in recent years mainly due to the concepts
borrowed  from the statistical mechanics of spin glasses, such as
the so-called  Random Energy Model (REM) first suggested by
Derrida \cite{Derrida}.

Noting  the properties of REM --- a rugged free energy landscape
and the statistical independence of states ---
Bryngelson and Wolynes first proposed the use of REM for
 protein folding
\cite{Wolynes}. Other approaches, which employed the sophisticated
machinery of replica mean field theory and began with an, although
simplified, microscopic model of random heteropolymer chains, also
led to REM-like conclusions \cite{IndepenInt}.  

However simple technically, REM and related ideas have proven to be
extremely fruitful in protein folding studies
\cite{IndepenInt,Stein,Others,ProteinRelated,Design}. 
It turned out so successful, that at present, it is often
considered  not as just a model, which may or may not be
applicable under different circumstances, but as a simple and
adequate universal language.  While 
REM seems like technical theoretical jargon, it is closely
related to the set of physical insights which is often used
without mentioning  (and perhaps without considering)  any
theoretical concepts at all (eg. Ref. \cite{denovo}).

The validity of REM was proven \cite{IndepenInt,ShakEnum} for
maximally compact heteropolymers with a Gaussian distribution of
monomer interaction energies.  On the other hand, there are many
observations that are hardly understandable using REM 
\cite{Beer}. 
 Thus, the goal of the present work is to scrutinize the real
strength and validity of REM in the context of heteropolymer
freezing and protein folding studies.
  We will show that REM is indeed often applicable, but it is far
from being applicable universally; its applicability is a highly
non-trivial property, and it fails or is at least questionable in
many cases.

\section{Qualitative Considerations}

Heteropolymer freezing is said to be due to
the frustrated interplay between three fundamental polymeric
elements:  sequence, set of conformations available for the given
chain, and the nature of interaction between monomer species. More
specifically, the freezing  of a polymer in a certain
conformation  is governed by the competition of two
conflicting factors.  First,  how well are the interactions
 adjusted in this conformation?  Clearly, this depends on
the character of interaction between monomers. A broad
distribution of the types of monomer species favors energetically
strong preferences between conformations. Second, how many
conformations exist in the vicinity of a given conformation?
 This defines the entropy and depends entirely on the
internal geometry of the space of available conformations 
(chain flexibility, topology, etc.).  

REM specifies this picture of freezing by the assertion that for each
sequence, the polymer energy  $E_{\alpha}$ of conformation
$\alpha$ is taken randomly from probability distribution over disorder
$P(E)$, which is the same for all $\alpha$; furthermore, they are
taken independently, such that joint probability is directly
expressed in terms of $P(E)$: $P(E_{\alpha}, E_{\beta}) =
P(E_{\alpha})P(E_{\beta})$.  Obviously, this cannot be valid
exactly.  Indeed, for conformation $\beta$ which is only a minor
conformational rearrangement of $\alpha$, many 
contributions to the respective energies are identical and thus the
energies are clearly correlated.  REM validity requires that such pairs
of conformations are rare.

The simplest quantitative measure of statistical dependence
between the energies of two given conformations $\alpha$ and $\beta$
over the set of sequences can be obtained by taking correlation
$\left< E_{\alpha} E_{\beta} \right> - \left< E_{\alpha} \right>
\left< E_{\beta} \right> = \left< \delta \! E_{\alpha} \delta \!
E_{\beta} \right>$, where $\left< \ldots \right>$ denotes averaging
over sequences.   REM invalidity can be demonstrated by the
non-vanishing of this correlator. We consider the Hamiltonian 
\begin{equation}
{\cal H} = \sum_{I \neq J} B(s_I,s_J) \ \ 
\delta \left(  {\bf r}_I  - {\bf r}_J  \right)
\label{eq:Hamiltonian}
\end{equation}
where $s_I \in \{1,\ldots,q\}$ is the species of monomer number
$I$, $q$ is the number of species,
$B_{ij}$ is the matrix  of species-species interactions, ${\bf r}_I$
is the position of monomer $I$, and $\delta \left(  {\bf r}  - {\bf
r}^{\prime}  \right)$ is the function that is concentrated on
nearest neighboring space points.  We implicitly assume here, that
position vectors ${\bf r}_I$ are such that the conditions of chain
connectivity (unitary spatial distance between sequential monomers),
excluded volume and constant density (on the lattice, for instance,
every site is occupied by one and only one monomer) are all met. 
This is the most general model for the case when heterogeneity comes
exclusively from pairwise, short range interactions \cite{Imprint}. 
For the Hamiltonian (\ref{eq:Hamiltonian}), we find
\begin{equation}
\left< \delta \! E_{\alpha} \delta \! E_{\beta} \right> =  
\left< \delta \! B ^2\right> {\cal Q} _{\alpha , \beta} + N z^2
\sum_{ijk}p_i  \delta \! B_{ij}   p_j  \delta \! B_{jk}  p_k 
\label{eq:Ecorr}
\end{equation} 
where $p_i$ is the probability of finding a monomer of species $i$,
$\delta \! B_{ij} \equiv B_{ij} - \sum_{kl} p_k B_{kl} p_l $, 
$\left< \delta \! B ^2\right> \equiv \sum_{ij} p_i \left( \delta \!
B_{ij} \right) ^2 p_j$ is the variance of the elements of the
interaction matrix, and ${\cal Q}_{\alpha
\beta} \equiv \sum_{I \neq J} 
\delta({\bf r}_I^\alpha - {\bf r}_J^\alpha) 
\delta({\bf r}_I^\beta  - {\bf r}_J^\beta)$ is the conventionally
defined overlap between conformations \cite{Others,ShakEnum}.  We have
taken into account here the condition that the polymer is maximally
compact, so that each monomer has $z$ space neighbors, and thus
$\sum_{I \neq J \neq K} \delta({\bf r}_I^\alpha - {\bf r}_J^\alpha) 
\delta({\bf r}_J^\beta  - {\bf r}_K^\beta) = Nz^2$. 
Thus, we see that there are two contributions to the correlator,
one dependent on conformations and the other independent. 

Eq.~(\ref{eq:Ecorr}) formally shows how aspects of interactions
($B_{ij}$), conformation space (${\cal Q}_{\alpha \beta}$), and 
sequences ($p_i$) enter into the nature of  statistical dependence.
In the following section, we use eq.~(\ref{eq:Ecorr}) to
demonstrate deviations from REM induced by manipulating each of
these three elements.

\section{Deviations from REM}

\subsection{Conformations}

The first term of the
energy correlator eq.~(\ref{eq:Ecorr})  is proportional to the
number of bonds that conformations $\alpha$ and
$\beta$ have in common, ${\cal Q}_{\alpha \beta}$; clearly,
identical bonds give identical contributions to energies of both
conformations, thus yielding a dependence.  
However, if large ${\cal Q}_{\alpha \beta}$ are rare, then REM
remains a good approximation.  To examine this for simple
computational models typically employed, we have
computed $P({\cal Q}) = \sum_{\alpha \beta}
\delta({\cal Q}_{\alpha \beta} - {\cal Q})$ for a variety of
conformational spaces.  

One type of space consists of maximally
compact conformations (27-mers and 36-mers in $3 \times 3
\times 3$ and $3 \times 3 \times 4$ respectively) which are typically
used in freezing simulations \cite{Others,ShakEnum,Enum}; maximally
compact conformations of relatively short chains favor REM validity as
small rearrangements are not possible and $P({\cal Q})$ will have
virtually no contributions in the large ${\cal Q}$ range.  
In comparison, we considered
maximally compact crumpled 64-mer conformations on the  $4 \times 4
\times 4$ lattice.  In crumpled conformations, neighbors along the
chain are likely to be neighbors in space \cite{Crumpled}.  In this
sense, crumpled conformations
 perhaps crudely model the effect of secondary structure.
In our lattice model, the $4 \times 4 \times 4$ cube can be broken down
into eight $2 \times 2 \times 2$ subcubes, and 
crumpled conformations  fill every site in a given subcube before
entering a new one.

\begin{figure}[tb]
\epsfxsize=3in
\centerline{\epsplace{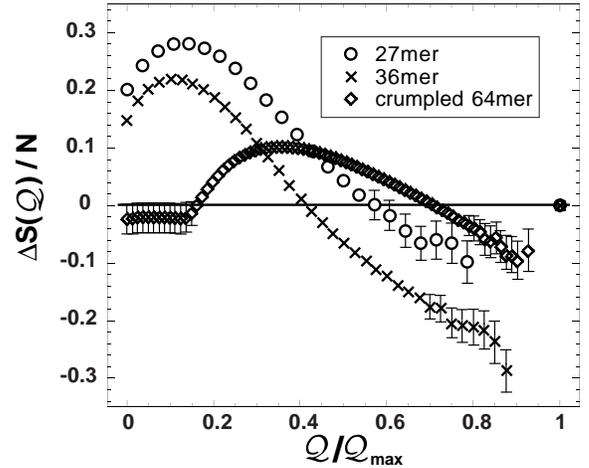}}
\bigskip
\caption{
$\Delta S({\cal Q}) \equiv \ln [P({\cal Q})/
P({\cal Q}={\cal Q}_{\rm max})]$ for compact
27-mers, compact 36-mers, and compact \& crumpled 64-mers.  
The discrete region boundary  varies greatly: ${\cal Q}_d/{\cal
Q}_{\rm max}
\approx 0.6,0.4,0.7$ for 27-mers, 36-mers, and crumpled 64-mers
respectively.
 }
\label{fig:SQ}
\end{figure}

\begin{figure}[bt]
\epsfxsize=3in
\centerline{\epsplace{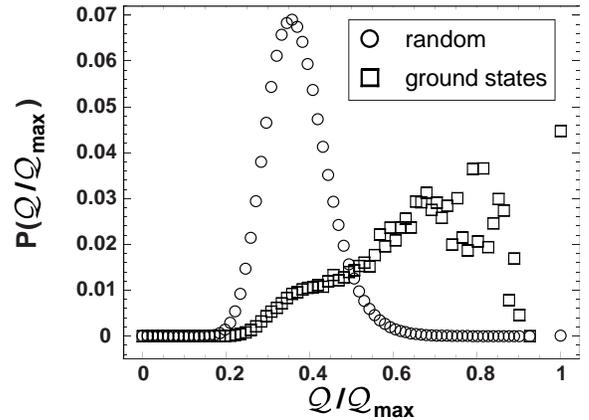}}
\bigskip
\caption{
$P({\cal Q}/{\cal Q}_{\rm max})$ for  crumpled, compact  64-mers.
We compare the results of conformations taken at random with that
obtained by comparing the degenerate ground state conformations
of  1000 sequences with Ising interactions.
Overlapping ground states  signal a  departure from REM. 
 }
\label{fig:PQfrac}
\end{figure}

As we see in Fig.~1, $P({\cal Q})$ for all conformational spaces studied
are peaked at small ${\cal Q}$ (this peak is not at
${\cal Q}=0$  due to finite size effects).  
As ${\cal Q}$ increases,
$P({\cal Q})$ decreases exponentially, and above a particular 
${\cal Q}_d$, 
 there are at most ${\cal O}(1)$ conformations
available.  Thus, for a space with small ${\cal Q}_d$,
there are few states with large overlap and REM is favored.
In Fig.~1, we see that crumpled conformations have the greatest
${\cal Q}_d$ as they allow a greater possibility of rearrangements
on small scales (large ${\cal Q}$).

With the knowledge of the relative favorability of a given
conformation space for REM validity, what is the effect of
these statistical dependencies on the nature of freezing? 
While REM formally means that all states are statistically
independent, what is relevant is the low energy states.
For example, degenerate ground states (as is common in
models with ``discrete'' interactions \cite{DREM}) 
will not overlap if they are statistically independent.  
This holds for 27-mers and 36-mers with Potts
interactions ($B_{ij} =\delta_{ij}$), whose ground states 
yield a $P({\cal Q})$ which is indistinguishable from
that of conformations taken at random  \cite{ShakEnum,DREM}.
In the light of our previous discussion, REM appears
to be valid because ${\cal Q}_d$ is sufficiently small.
The situation is different for crumpled 64-mers:
upon enumerating \cite{Enum}
the energies of all conformations for 1000
sequences with Ising (2 letter Potts) interactions 
and comparing the ground states (Fig.~2), we see that the increase in
${\cal Q}_d$ for crumpled 64-mers is sufficient such that  REM fails for
this conformational space.  
This does not merely demonstrate that REM breaks for crumpled 64-mers in
particular, but rather REM validity is clearly not an a priori property
of all conformation spaces in general, even in three dimensions
\cite{IndepenInt}.

\subsection{Interactions}

Interactions enter into the conformational dependent term of 
eq.~(\ref{eq:Ecorr}) through the variance of the elements of the
interaction matrix, and thus this term is present for all types of
interactions.   
However, interactions play a more dramatic role in the conformational
independent term in eq.~(\ref{eq:Ecorr}), as it vanishes for many
models, but not, for example, if there is one monomer species that
interacts particularly strong with all others; in this case, there is a
correlated contribution even when there is no bond in common.   The
appearance of the conformation independent term 
 signals a departure from REM, as even states with vanishing ${\cal Q}$
are statistically dependent.

What forms of interactions have this residual (conformation
independent) statistical dependence?  One notable example is the HP
model   ($B_{ij} = \delta_{i1}\delta_{1j}$) \cite{Others}.
For even composition overall ($p_i=1/q$), but not fixed for a 
particular sequence, the conformation independent term does not vanish.
%
%
However,
for composition fixed to be even for each sequence and finite chains,
there is an additional term to eq.~(\ref{eq:Ecorr}) which leads to a
{\em negative \/} conformation independent constant. 
Thus, typical conformations will have a
small overlap, but the effect of this overlap is canceled by the
negative constant and the result is an effective statistical
independence for typical conformations (but anti-dependence
(dependence) for conformations with smaller (larger) overlap). 
Thus, REM may appear to be a good approximation, but only due to a
complicated cancellation of factors.  
Upon examination with eq.~(\ref{eq:Ecorr}), the Miyazawa and Jernigan
matrix (MJM) \cite{MJ}, a set of amino acid interaction potentials
derived from protein statistics, seems like the HP model plus some
noise; thus our analysis for HP is applicable to MJM.
For models with an even composition and a  symmetric contribution from
monomer species, such as Potts  or Independent Interaction Model (IIM),
in which $B_{ij}$ are taken from a Gaussian distribution
\cite{IndepenInt} interactions, the conformation independent term in
eq.~(\ref{eq:Ecorr}) vanishes and the only statistical dependence comes
from conformational overlap.

\subsection{Sequences}

An uneven composition also leads to the non-vanishing of the
conformation independent term (and thus a 
a statistical dependence) in many models,
including Potts and IIM, since there is an imbalance as in the HP
model discussed above.  

However, while we have been speaking about random sequences, protein
sequences are not random. Indeed, theoretically,  
models of ``minimal frustration'' \cite{Wolynes},
``sequence selection'' \cite{Design},  or ``imprinting''
\cite{Imprint} in which sequences are chosen  to have low energies in a
desired target conformation $\star$,   have been employed as a better
model of proteins.    If states were truly statistically independent, 
these procedures would act to ``pull down'' the energy of the ground
state alone; however,  statistical dependence between states means
that the energy of other states are pulled down as well.
Moreover, sequence selection has also been experimentally realized in  
{\it de novo} protein design \cite{denovo}, and REM
 is an  implicit assumption in such experiments: if
design leads to many states (not just the desired state $\star$) with
low energy, this procedure  fails. 

For our purpose of describing the validity of REM, 
sequence selection is also a useful
tool to demonstrate  statistical dependence.
In this sense, selection acts as a ``field'' 
in which we test the response of the energies of other conformations 
to the manipulation of the energy of
$\star$. In Fig.~3, we see from the density of states, 
$P(E,{\cal Q}) \equiv 
\sum_{\alpha} \delta \!\left( {\cal Q} - {\cal Q}_{\alpha \star} \right)
 \delta \!\left( E - {\cal H}_\alpha \right)$,
of a well designed sequence with Ising interactions that 
 sequence selection acts on
all states $\alpha$ 
to a degree that is roughly linear in ${\cal Q}_{\alpha \star}$. 
Therefore, the energy of a particular conformation
$\alpha$ is largely determined by  ${\cal Q}_{\alpha \star}$.  
Thus, we  see that  
sequence selection does not pull down the energy of a desired
conformation but rather affects the whole density of states.  
This is a dramatic deviation from REM.

However, this effect is not necessarily detrimental.  
For models with a degree of statistical anti-dependence (eg.
fixed, even composition HP), selection may act to push up the energy
levels of conformations other than $\star$.  Moreover, as this
statistical anti-dependence varies between sequences, it
could be considered as a sequence selection criteria (in additional to
optimization of the ground state) and therefore is a possible means to
``design out'' unwanted conformations \cite{Beer}.
 Also, recent models of ``protein folding funnels'' \cite{Others}
require conformational space deviations from REM: without some
relationship between ${\cal Q}$ and energy, there would be no ``path''
to the ground state and kinetics would be simply a random search
through conformation space, requiring the Levinthal time 
\cite{REMKinetics}.

\begin{figure}[bt]
\epsfxsize=3in
\centerline{\epsplace{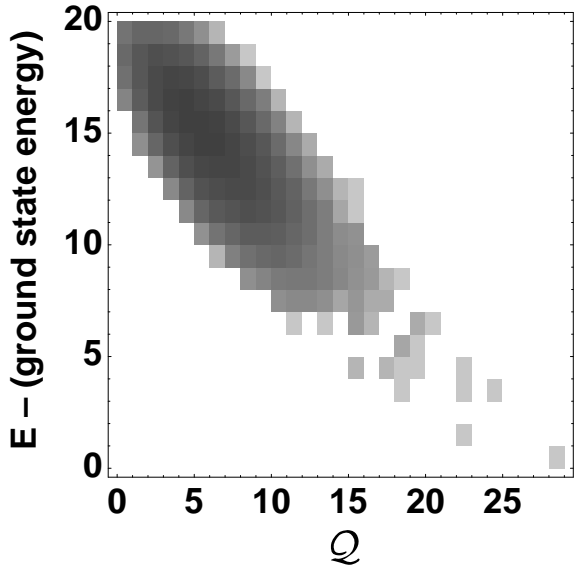}}
\bigskip
\caption{ 
$P(E-E_{\rm gnd},{\cal Q})$ 
for maximally compact
27-mer  conformations (${\cal Q}_{\rm max}=28$), Ising interactions
($B_{ij} = \delta_{ij}$), and a sequence chosen to have a low
ground state energy.
 }
\label{fig:DOS}
\end{figure}

\section{Conclusions}

While our
discussion employs simplified models, such as lattice conformation
spaces and a Hamiltonian~(\ref{eq:Hamiltonian}) which includes only
pairwise interactions, 
the aspects which push the system away from REM validity are
model independent.
Is heteropolymer freezing  well described by REM?  This complicated
question is clearly dependent on all aspects of the system under
question.  What is clear is that REM validity is not guaranteed for
all heteropolymers. For the models
commonly employed, it is notable that HP and MJM have extremely
questionable REM validity.

Is protein folding well described by REM?  
From the point of view of eq.~(\ref{eq:Ecorr}), 
interactions between amino acids are in the
HP/MJM class; biologically, the possibility of statistical
anti-dependence is quite intriguing, and perhaps sheds light on
the evolutionary significance of the nature of amino acid
interactions. 
As for the nature of the space of protein
conformations, one must consider possible kinetic restrictions as
well as the role of secondary structure, which highly restricts the
nature of conformational space, perhaps hindering REM.  
Thus, with possible deviations from REM arising from both
interactions and conformational space, REM validity for proteins is
questionable.

\bigskip

\bigskip

\centerline{\bf ACKNOWLEDGEMENTS}

\medskip

We have  benefited from discussions with 
A.  Gutin,  M. Kardar, and E. Shakhnovich. 
The work was supported by NSF (DMR 90-22933)
and Project SCOUT (ARPA  MDA972-92-J-1032).
AYG  acknowledges the support of Kao Fellowship.


\begin{thebibliography}
\medskip
%
\bibitem{Derrida} B. Derrida, {\em Phys. Rev. Lett} {\bf 45}
(1980) 79.
%
\bibitem{Wolynes} J. D. Bryngelson and P. G. Wolynes, {\em Proc.
Nat. Acad. Sci., USA} {\bf 84} (1987) 7524.
%
\bibitem{IndepenInt} E. Shakhnovich and   A. Gutin, {\em Biophys.
Chem.} {\bf 34}, 187 (1989); {\em J. Phys.} {\bf A22}, 1647 (1989).
%
%
\bibitem{Stein} P.G. Wolynes, in {\em Spin Glass Ideas in
Biology}, edited by D. Stein (World Scientific, Singapore, 1991).
%
\bibitem{Others}
 K. Yue and  K. A. Dill, {\em Proc. Nat. Acad. Sci., USA} {\bf
89}, 4163 (1992); N. D. Socci and J. N. Onuchic, {\em J. Chem.
Phys.}, {\bf 101}, 1519 (1994); J. D. Bryngelson, J. N. Onuchic,
N. D. Socci, and P. G. Wolynes, 
    {\em Proteins}, {\bf 21}, 167 (1995).
%
\bibitem{ProteinRelated} E. I. Shakhnovich and A. M. Gutin, {\em
J. Theor. Biol} {\bf 149},  537 (1991); J. D. Bryngelson, {\em J.
Chem. Phys.} {\bf 100}, 6038 (1994);  V. S. Pande, A. Yu.
Grosberg, and T. Tanaka, {\em J. Chem. Phys.} {\bf 103}, 9482 (1995).
%
%
\bibitem{Design} E.I. Shakhnovich  and A.M. Gutin, 
{\em Proc. Nat. Acad. Sci., USA}  {\bf 90}, 7195 (1993).
%
\bibitem{Imprint} V. S. Pande, A. Yu. Grosberg, and
T. Tanaka,  {\em Macromolecules} {\bf 28}, 2218  (1995).
%
\bibitem{denovo} M. H. Hecht, J. S. Richardson, and D. C. Richardson,
{\em Science} {\bf 249}, 884 (1990).
%
\bibitem{ShakEnum} E. Shakhnovich,  and A. Gutin, {\em J Chem.
Phys.}, {\bf 93}, 5967 (1990).
%
\bibitem{Beer} K. Yue, K. M. Fiebig, P. D. Thomas, H. S. Chan, K.
A. Dill, and  E. I. Shakhnovich, {\em Proc. Nat. Acad. Sci, USA}
{\bf 92}, 325 (1995);
H. Li, R. Helling, C. Tang, and N. Wingreen, {\em cond-mat preprint}
{\bf 9603016} (1996).
%
\bibitem{Enum} V. S. Pande, C. Joerg, A. Yu. Grosberg, and T.
Tanaka, {\em J. Phys.} {\bf A27}, 6231 (1994).
%
\bibitem{Crumpled} A. Yu. Grosberg, S. K. Nechaev, 
and E. I. Shakhnovich, 
{\em J. de Phys. (France)} {\bf 49}, 2095 (1988).
%
\bibitem{DREM} A. M. Gutin and E. I. Shakhnovich, {\em J
Chem. Phys.}, {\bf 98}, 8174 (1993).
%
%
\bibitem{MJ} S. Miyazawa and R. Jernigan, {\em Macromolecules} {\bf
18},  534 (1985).
%
\bibitem{REMKinetics}  J. D. Bryngelson and P. G. Wolynes, 
{\em J. of Phys. Chem.} {\bf 93} 6902 (1989).
%
\end{thebibliography}
\end{document}